\documentclass[%
 reprint,
 superscriptaddress,
 amsmath,amssymb,
 aps,
]{revtex4-2}

\usepackage{graphicx}
\usepackage{hyperref}
\hypersetup{colorlinks,urlcolor=blue, linkcolor=blue}
\begin{document}

\title{Room-temperature optically detected coherent control of molecular spins}

\author{Adrian Mena}\thanks{These authors contributed equally to this work.}
\affiliation{James Watt School of Engineering, University of Glasgow, Glasgow, G12 8QQ, UK.}
\affiliation{ARC Centre of Excellence in Exciton Science, School of Physics, UNSW Sydney, Sydney, 2052, NSW, Australia.}

\author{Sarah K. Mann}\thanks{These authors contributed equally to this work.}
\affiliation{James Watt School of Engineering, University of Glasgow, Glasgow, G12 8QQ, UK.}

\author{Angus Cowley-Semple}\thanks{These authors contributed equally to this work.}
\affiliation{James Watt School of Engineering, University of Glasgow, Glasgow, G12 8QQ, UK.}

\author{Emma Bryan}
\affiliation{Department of Materials and London Centre for Nanotechnology, Imperial College London, Prince Consort Road, London, SW7 2AZ, UK.}

\author{Sandrine Heutz}
\affiliation{Department of Materials and London Centre for Nanotechnology, Imperial College London, Prince Consort Road, London, SW7 2AZ, UK.}

\author{Dane R. McCamey}
\affiliation{ARC Centre of Excellence in Exciton Science, School of Physics, UNSW Sydney, Sydney, 2052, NSW, Australia.}

\author{Max Attwood}
\email{m.attwood@imperial.ac.uk}
\affiliation{Department of Materials and London Centre for Nanotechnology, Imperial College London, Prince Consort Road, London, SW7 2AZ, UK.}

\author{Sam L. Bayliss}
\email{sam.bayliss@glasgow.ac.uk}
\affiliation{James Watt School of Engineering, University of Glasgow, Glasgow, G12 8QQ, UK.}

\begin{abstract}Benefiting from both molecular tunability and versatile methods for deployment, optically interfaced molecular spins are a promising platform for quantum technologies such as sensing and imaging. Room-temperature optically detected coherent spin control is a key enabler for many applications, combining sensitive readout, versatile spin manipulation, and ambient operation. Here we demonstrate such functionality in a molecular spin system. Using the photoexcited triplet state of organic chromophores (pentacene doped in a para-terphenyl host), we optically detect coherent spin manipulation with photoluminescence contrasts exceeding 10\% and microsecond coherence times at room temperature. We further demonstrate how coherent control of multiple triplet sublevels can significantly enhance optical spin contrast, and extend optically detected coherent control to a thermally evaporated thin film, retaining high photoluminescence contrast and coherence times of order one microsecond. These results open opportunities for room-temperature quantum technologies that can be systematically tailored through synthetic chemistry.
\end{abstract}

\maketitle

Optically addressable electronic spin systems that can operate effectively under ambient conditions have emerged as a promising platform for applications such as quantum sensing and imaging, with examples including solid-state defects in diamond \cite{Taylor2008, schirhagl2014nitrogen}, silicon carbide \cite{Koehl2011}, and 2D materials \cite{Gottscholl2020}. By harnessing the unique properties of quantum mechanics, such systems hold promise to extend the capabilities of conventional sensing/imaging. For example, room-temperature optically detected coherent control of electron spins offers advanced detection of quantities ranging from magnetic and electric fields \cite{Taylor2008, Wolf2015, Dolde2011} to temperature \cite{Kucsko2013, toyli2013fluorescence}. 

Complementary to solid-state defects, electronic spins in molecules are promising systems to realise spin-light interfaces due to their chemical versatility. For example, ground-state spins in organometallic molecules can offer analogous functionality to solid-state defects in a synthetically tunable platform \cite{bayliss2020}, and photoexcited molecular spins can be deployed in rich applications spanning structure determination of biomolecules \cite{valentin2014, hintze2016laser, Bertran2021}, electrical spin readout \cite{mccamey2008}, multi-spin coupling \cite{Gorgon2023}, and entanglement generation \cite{Filidou2012, rugg2019photodriven}. However, an outstanding challenge is to achieve optically detected coherent control of molecular spins at room temperature.

Organic chromophores with photoexcited electronic spins are an attractive system in which to realise room-temperature optically detected coherent control due to their bright optical transitions, weak spin-orbit coupling, and ability to be integrated with a range of targets. Such functionality would open a range of opportunities. In molecular biology, adding the high sensitivity of optical detection to recently developed spin labelling techniques \cite{valentin2014, hintze2016laser, Bertran2021} could enable structural analysis of individual biomolecules \cite{Shi2015}. Conjugating chromophores with an optically accessible spin to biological targets could realise spin-enhanced molecular labels for ultrasensitive fluorescent assays \cite{Miller2020}, or for use as contrast agents for biomedical imaging in scattering media \cite{Hegyi2013}. The versatile deposition techniques available to organic chromophores could further be used to maintain proximity between sensor and target, even on textured or flexible surfaces, for applications such as wide-field spin-based imaging with high spatial resolution \cite{Chen2013}. 

As a platform for the above opportunities, here we demonstrate optically detected coherent control of molecular spins at room temperature. We achieve this using the photoexcited triplet states of an organic chromophore---pentacene, doped in a para (\textit{p})-terphenyl host---using its spin-selective intersystem crossing rates that allow for both optical initialisation and readout at room temperature. Pentacene in \textit{p}-terphenyl has been a powerful system for molecular spin-optical functionality, being used for the first demonstration of single spin magnetic resonance (at cryogenic temperatures) \cite{Kohler1993, Wrachtrup1993}, and subsequently finding diverse applications including in dynamic nuclear polarisation \cite{Tateishi2014}, and room-temperature masing \cite{Oxborrow2012}. Recently, continuous-wave optically detected magnetic resonance was reported for pentacene doped in a picene host at room temperature \cite{Moro2022}, while the versatility of the acene family (to which pentacene belongs) is highlighted by their ability to be covalently immobilized as monolayers on surfaces \cite{Zholdassov2023}, or used as dyes for biomedical imaging \cite{Pansare2014}. Here we demonstrate Rabi oscillations with photoluminescence contrasts exceeding 10\%, microsecond spin coherence times ($T_2$), and an ensemble dephasing time ($T_2^\star$) of 390\,ns under ambient conditions with a pentacene:\textit{p}-terphenyl crystal. We then demonstrate coherent control of multiple spin transitions as a way of significantly enhancing spin-optical contrast and spin dynamics. Finally, we highlight the versatile deposition capabilities available to molecular spin systems by implementing optical readout of coherent control in a 100\,nm thermally evaporated thin film. Overall, these results demonstrate the potential for room-temperature quantum sensing and imaging with a molecular platform.

\section*{Room-temperature optical interface to pentacene spins}\label{cw-ODMR}

\begin{figure}[t]
    \centering
    \includegraphics[width=\columnwidth]{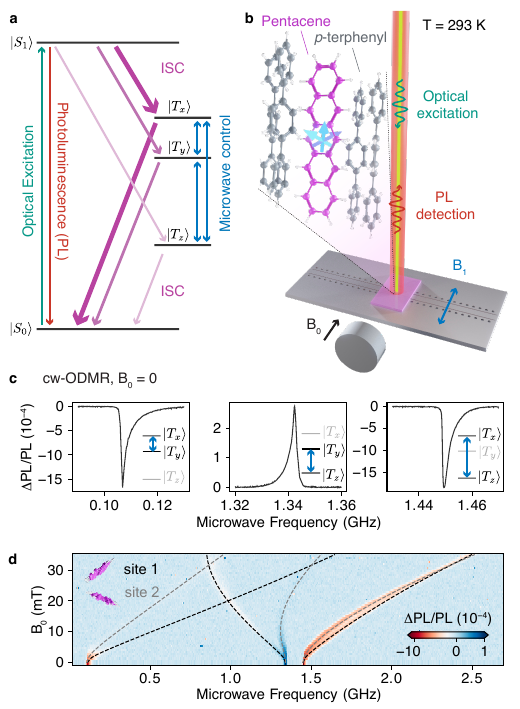}
    \caption{\textbf{Room-temperature spin-optical interface in pentacene:p-terphenyl} \textbf{a,} Simplified energy-level diagram for pentacene illustrating the processes involved in spin-polarised triplet formation, microwave spin control, and optical spin detection. \textbf{b,} Schematic illustration of the experimental setup for spin initialisation, manipulation, and readout. \textbf{c, d,} Continuous-wave ODMR spectra of a 0.01\% doped pentacene:\textit{p}-terphenyl single crystal at \textbf{c}, zero-field and \textbf{d}, as a function of magnetic field. Simulations are shown by dashed lines (see Supplementary Text S2 for simulation details).}
    \label{fig1}
\end{figure}

\begin{figure*}[htb!]
    \centering
    \includegraphics[width=\textwidth]{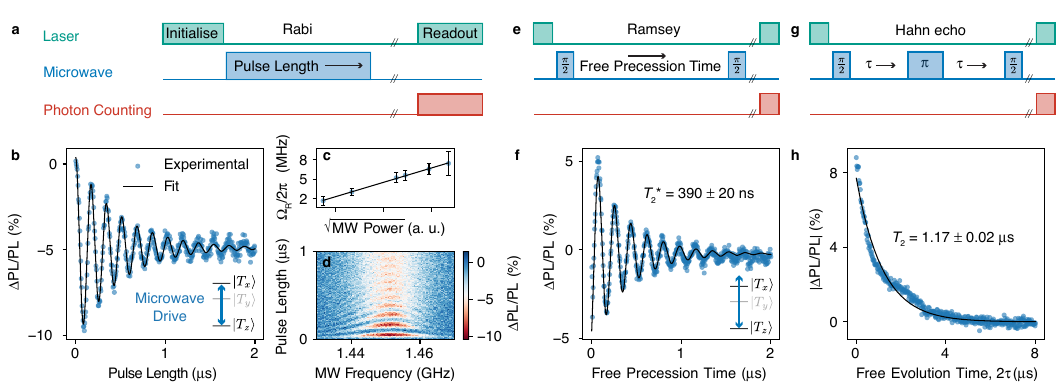}
    \caption{\textbf{Optically detected coherent control of molecular spins at room temperature.} (All plots are for driving the ${|T_x\rangle \leftrightarrow |T_z\rangle}$ transition.)  \textbf{a,} Pulse sequence for the Rabi experiment. \textbf{b,} Rabi oscillations. \textbf{c,} Microwave-power dependence of the Rabi oscillation frequency, $\Omega_R$. \textbf{d,} Rabi chevron plot showing the PL contrast as a function of microwave frequency and pulse length. \textbf{e,} Pulse sequence for the Ramsey experiment. \textbf{f,} Ramsey trace (measured with a detuning of 5\,MHz) yielding \(T_2^\star\,=\,390\,\pm\,20\,\)ns.  \textbf{g,} Pulse sequence for the Hahn-echo experiment. \textbf{h,} Hahn-echo measurement yielding $T_2=1.17 \pm 0.02$\,$\mu$s. Fits for all traces are shown with black lines (see Supplementary Text S3 for details).} 
    \label{fig2}
\end{figure*}

Fig. \ref{fig1}a shows a simplified energy level diagram illustrating the key components of pentacene's optical-spin interface (see Supplementary Text S1 for further details). Following optical excitation of pentacene molecules from the singlet ground state, $|S_0\rangle$, to the singlet excited state, $|S_1\rangle$, they can either decay back to $|S_0\rangle$, and emit photoluminescence (PL), or undergo inter-system crossing (ISC) to populate the long-lived (depopulation rates ranging from $\sim$35-500\,$\mu$s \cite{Wu2019}) spin-1 triplet state with an overall yield of $\simeq$63\% \cite{Takeda2002}. The zero-field splitting removes the degeneracy of the triplet sublevels at zero magnetic field, giving rise to three sublevels, $|T_x\rangle$, $|T_y\rangle$ and $|T_z\rangle$. (The pentacene molecular $X$, $Y$, and $Z$ axes are the molecule's long, short, and out-of-plane axes, respectively.) The zero-field splitting parameters, $D = 1396$\,MHz and $E = -53$\,MHz \cite{Yang2000}, give transition frequencies between the sublevels of 107\,MHz ($|T_x\rangle \leftrightarrow |T_y\rangle$), 1.34\,GHz ($|T_y\rangle \leftrightarrow |T_z\rangle$) and 1.45 GHz ($|T_x\rangle \leftrightarrow |T_z\rangle$). Anisotropic ISC depopulation rates of the triplet sublevels to $|S_0\rangle$ result in spin-dependent triplet sublevel lifetimes \cite{Wu2019}. Irradiation with microwaves resonant with the triplet sublevels' transition frequencies affects the photoluminescence intensity by transferring populations between longer- and shorter-lived sublevels. This in turn affects the repopulation rate of the $|S_0\rangle$ ground state, and therefore the ability of the molecules to be re-excited and emit PL, thereby enabling optical spin-state detection through optically detected magnetic resonance (ODMR) \cite{Kohler1993, Wrachtrup1993}.

Fig. \ref{fig1}b shows the key components of our experimental setup (see Methods and Supplementary Fig. S3 for further details). We achieve spin control from the microwave field ($B_1$) applied by a coplanar waveguide. The samples is excited with a laser (520\,nm) and we detect the red-shifted photoluminescence ($>$600\,nm) for spin-dependent readout. Zeeman splitting can be induced with a static magnetic field ($B_0$, provided by a permanent magnet).

As a pre-requisite for optically detected coherent control, we first measure continuous-wave (cw) ODMR of a single crystal of pentacene doped at 0.01\% into \textit{p}-terphenyl (Fig. \ref{fig1}c). For the $|T_x\rangle \leftrightarrow |T_y\rangle$ and $|T_x\rangle \leftrightarrow |T_z\rangle$ transitions, we observe an ODMR contrast of approximately $-0.2\%$, while for the $|T_y\rangle \leftrightarrow |T_z\rangle$ we observe a positive contrast, an order of magnitude smaller, of approximately 0.02\% (see Supplementary Text S2 for further discussion). The asymmetric ODMR lineshapes arise from second-order hyperfine interactions between the triplet spins and the pentacene protons \cite{Kohler1999}. Fig. \ref{fig1}d shows the cw-ODMR spectra as a function of an external magnetic field where the transition frequencies shift as a result of Zeeman splitting (the magnetic field was not specifically aligned to the crystal axes). (We note that due to the two inequivalent pentacene lattice sites in \textit{p}-terphenyl, we resolve two sets of resonances for each transition, and the changes in PL contrast with field arise due to the field-induced hybridisation between zero-field sublevels.)

\section*{Room temperature optically detected coherent control}\label{coherent_ODMR}

We next demonstrate room-temperature optically detected coherent control on the same sample, including Rabi oscillations, Ramsey interferometry and Hahn echoes. Each of these pulsed experiments follows the same basic sequence. First, we optically initialise the photoexcited triplet state with a laser pulse, resulting in triplet sublevel populations $P_x:P_y:P_z \simeq 0.76:0.16:0.08$ (see Supplementary Text S4 for further discussion) \cite{Sloop1981}. We then apply a microwave spin control sequence. For spin-dependent optical readout, we incorporate a delay time (tens of microseconds) following microwave manipulation to repopulate the ground state in a spin-dependent fashion due to the different depopulation rates from the triplet sublevels to the ground state, $|S_0\rangle$. Photoluminescence collection during a final read-out laser pulse probes the singlet ground state, completing the spin readout. All following experiments were performed at zero applied magnetic field.

We first focus on the $|T_x\rangle \leftrightarrow |T_z\rangle$ transition before discussing multi-level control involving the $|T_y\rangle \leftrightarrow |T_z\rangle$ transition. (Measurements on the $|T_x\rangle \leftrightarrow |T_y\rangle$ transition are included in the Supplementary Information; Fig. S4.) We drive Rabi oscillations using the pulse sequence in Fig. \ref{fig2}a, comprising a variable length microwave pulse. We observe a PL contrast of approximately 10\% (Fig. \ref{fig2}b), significantly higher than the cw-ODMR signals in Fig. \ref{fig1}. The Rabi oscillation frequency shows the expected square-root dependence on the microwave power (Fig. \ref{fig2}c, Supplementary Fig. S5), and the expected behaviour with microwave detuning (Fig. \ref{fig2}d, Supplementary Fig. S6). We next use a Ramsey sequence (Fig.\,\ref{fig2}e, 5\,MHz detuning) to measure the ensemble dephasing time, \(T_2^\star = 390\,\pm\,20\)\,ns (see Supplementary Fig. S7 for Ramsey data as a function of detuning). We next perform a Hahn-echo sequence, as depicted in Fig. \ref{fig2}f to measure the coherence time, $T_2$. We fit the data to an exponential decay to extract the coherence time, $T_2 = 1.17 \pm 0.02$\,$\mu$s (Fig. \ref{fig2}g). 

\section*{Coherent control of multiple triplet transitions}\label{multi-level}
In planar aromatic hydrocarbons such as pentacene, the triplet sublevels which are most rapidly populated through inter-system crossing are also typically the ones which are most rapidly depopulated, and similarly, the sublevel with the slowest depopulation rate also typically has the slowest population rate \cite{clarke2008,anthenius1974}. For example, in pentacene, the  $|T_x\rangle$ sublevel has both the fastest population rate and the fastest depopulation rate, while the $|T_z\rangle$ sublevel has the slowest population and depopulation rates. At first sight, this behaviour appears to present a trade-off between optimising triplet spin polarisation (determined by the fastest population rate), spin lifetimes (favouring slow depopulation rates), and fast experiment repetition rates (favouring fast depopulation rates). However, we can overcome this limitation by coherently controlling multiple triplet sublevel transitions. Specifically, the more highly populated, but shorter-lived $|T_x\rangle$ population can be transferred through a microwave pulse to the longer-lived levels---$|T_y\rangle$, $|T_z\rangle$---for control, and then subsequently transferred back to the more rapidly decaying $|T_x\rangle$ sublevel for effective optical readout. This multi-level control can therefore be used to overcome the above trade-off, enabling two long-lived, highly polarised triplet sublevels to be used for manipulation, while the third is capitalised on for efficient polarisation generation and optical readout. 

\begin{figure}[t]
    \centering
    \includegraphics[width=\columnwidth]{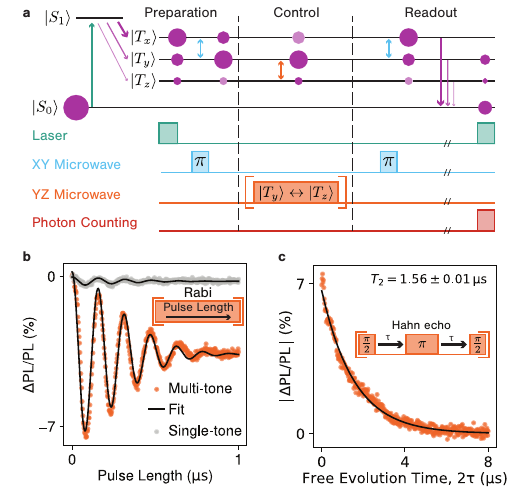}
    \caption{\textbf{Multi-level driving for enhanced optical readout of the ${|T_y\rangle \leftrightarrow |T_z\rangle}$ transition. a,} Multi-level control pulse sequence. A $\pi$-pulse on the ${|T_x\rangle \leftrightarrow |T_y\rangle}$ transition increases the spin polarisation between ${|T_y\rangle \leftrightarrow |T_z\rangle}$. A second microwave frequency controls the ${|T_y\rangle \leftrightarrow |T_z\rangle}$ transition, before population is transferred back to $|T_x\rangle$ through a final $\pi$-pulse for more effective optical readout. \textbf{b,} Rabi oscillations on the ${|T_y\rangle \leftrightarrow |T_z\rangle}$ transition measured with (orange) and without (grey) multi-level control. \textbf{c,} Hahn echo measured with multi-level control. Fits are shown with black lines.} 
    \label{fig3}
\end{figure}

To demonstrate this approach, we use the multi-level control scheme shown in Fig. \ref{fig3}a which consists of preparation, control, and readout sequences. During preparation, a laser pulse is followed by a microwave $\pi$-pulse on the $|T_x\rangle \leftrightarrow |T_y\rangle$ transition which transfers the initial (high) $|T_x\rangle$ population to the longer-lived $|T_y\rangle$ state. Following this, we use a second microwave tone to selectively address the $|T_y\rangle \leftrightarrow |T_z\rangle$ transition, whose levels have slower depopulation rates than $|T_x\rangle$. A final $\pi$-pulse on the $|T_x\rangle \leftrightarrow |T_y\rangle$ transition transfers the $|T_y\rangle$ population back onto the more quickly depopulating $|T_x\rangle$ level for more effective optical readout.

Figure \ref{fig3}b shows a Rabi oscillation on the $|T_y\rangle \leftrightarrow |T_z\rangle$ transition measured using this multi-level control approach. For comparison, we also show the signal for single-tone driving of the $|T_y\rangle \leftrightarrow |T_z\rangle$ transition (i.e., without using the $|T_x\rangle \leftrightarrow |T_y\rangle$ $\pi$-pulses such that the sequence is equivalent to the one shown in Fig. \ref{fig2}a). Multi-level control gives a PL contrast of 7\%, a nearly twenty-fold increase over single-tone driving, which gives a contrast of 0.4\% (see Supplementary Text S5 for further details). With this enhanced protocol in place, we next use it to measure a high-contrast Hahn echo on the $|T_y\rangle\leftrightarrow|T_z\rangle$ transition (Fig. \ref{fig3}c), yielding $T_2=1.56 \pm 0.01\,\mathrm{\mu}$s, similar to the coherence time of the $|T_x\rangle \leftrightarrow |T_z\rangle$ transition (Fig. \ref{fig2}h). Overall, this multi-level manipulation demonstrates how coherent control can be used to overcome the tradeoff between efficient optical spin readout and long spin lifetimes in organic triplets, and highlights how synthetic control of population/depopulation rates \cite{Ng2023} could be used to further increase optical spin contrast and experiment repetition rates, and therefore spin-detection signal-to-noise ratios.

\begin{figure*}[htb!]
    \centering
    \includegraphics[width=\textwidth]{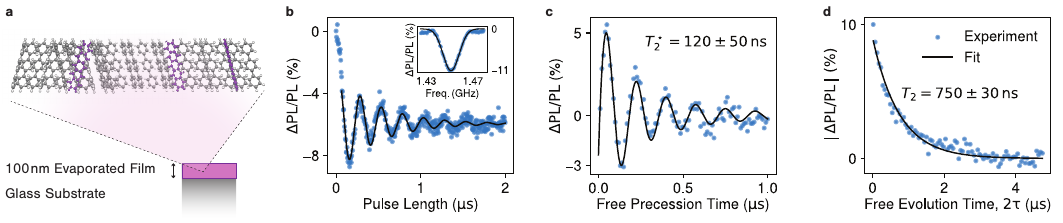}
    \caption{\textbf{Optically detected coherent control of a pentacene:\textit{p}-terphenyl thin film.} \textbf{a,} Schematic of the thin film (0.1\% doping concentration) and molecular ordering. \textbf{b,} Optically detected Rabi oscillations, and pulsed-ODMR spectrum (inset). \textbf{c, d,} Ramsey (5\,MHz detuning) and Hahn echo traces. Fits for all traces are shown in black lines.}
    \label{fig4}
\end{figure*}

\section*{Coherent manipulation in a thin film}\label{thin-film}
We next demonstrate the opportunities of molecular deposition techniques through room temperature optically detected coherent control of a thermally evaporated pentacene:\textit{p}-terphenyl thin film. Such films could open up novel opportunities for spin-based quantum sensors due to their ability to be grown with controlled thickness at few-nanometre precision \cite{Kang2014}; deposited on various surfaces, including flexible substrates \cite{kim2020}; and deployed with higher doping concentration ($\sim1\%$) than crystals \cite{Cui2020} whilst retaining molecular ordering---with pentacene's long axis preferentially aligning nearly perpendicular to the substrate \cite{Lubert-Perquel2018} (Fig. \ref{fig4}a). In the following, we demonstrate room-temperature optically detected coherent control of a 100 nm film (evaporated onto a glass substrate) with 0.1\% pentacene concentration (see Methods for thin-film growth details).

Figure \ref{fig4}b shows Rabi oscillations between the $|T_x \rangle\leftrightarrow |T_z \rangle$ sublevels of the thin film, demonstrating that this morphology retains high-contrast coherent control. Fig. \ref{fig4}b, inset shows the pulsed-ODMR spectrum (measured at a higher microwave drive strength than the Rabi oscillation) with a contrast reaching 11\%, further highlighting the preservation of high contrast in the film. Figures \ref{fig4}c, d show Ramsey and Hahn-echo measurements on the same film, yielding $T_2^\star=120\,\pm50\,\text{ns}$ and a $T_2=750\pm30\,\text{ns}$, comparable to the crystal measurements (Fig. \ref{fig2}), despite the order of magnitude higher doping concentration and distinct deposition method.

\section*{Sensing metrics}\label{sensitivity}
Our results demonstrate ingredients for using molecular spins for room-temperature quantum sensing. For such applications, key parameters are the sensing field that the spins can access (determined by the proximity between sensor and target) and their sensitivity to the relevant external physical quantity. Pentacene has a non-zero transverse zero-field splitting, making it first-order insensitive to magnetic fields close to zero field. However, we can use our results to estimate a magnetic-field sensitivity assuming a linear Zeeman splitting, which could be realised with a different molecule at zero field, or by applying an external field bias. For ensemble-based sensing, the relevant metric is the volume-normalised sensitivity, \(\eta^V\) \,\cite{Taylor2008,Zhou2020}. This quantity (equal to \(\eta\sqrt{V}\), where \(\eta\) is the bare sensitivity and \(V\) the collection volume) is independent of the collection volume and therefore allows direct comparison between experiments probing different volumes of spins. Using the experimental thin-film $T_2^\star=120$\,ns, pentacene doping concentration \(c_S=0.1\%\), measurement overhead time $t_\text{overhead}\simeq$350 $\mu$s (which accounts for initialisation, readout, and reset), optical spin-contrast $C=0.05$, and an estimated average number of photons per molecule per readout of \(n_\text{avg}=\)\(10^{-3}\) (reflecting imperfect photon collection efficiencies\,\cite{Taylor2008}) we calculate a DC volume-normalised sensitivity of \(\eta_{\text{DC}}^V\simeq800\,\text{nT}\,\mu\text{m}^{3/2}\,\text{Hz}^{-1/2}\) for the thin film, with similar values for the crystal (see Methods and Supplementary Table S1 for further details). Likewise, using the experimental \(T_2\) for the thin film, we calculate an AC volume-normalised sensitivity \(\eta_{\text{AC}}^V\simeq200\,\text{nT}\,\mu\text{m}^{3/2}\,\text{Hz}^{-1/2}\). While there are multiple routes to optimise these numbers further, as discussed below, they are promising given state-of-the art numbers for solid-state defects: \(\eta_\text{DC}^V = 34\,\text{nT}\,\mu\text{m}^{3/2}\,\text{Hz}^{-1/2}\)\,\cite{Barry_2016} and \(\eta_\text{AC}^V = 13\,\text{nT}\,\mu\text{m}^{3/2}\,\text{Hz}^{-1/2}\)\,\cite{Zhou2020, Wolf2015}. Considering the sensitivity scales as \(\eta^V\propto\frac{1}{C\sqrt{c_{s}}}\frac{\sqrt{t_{\text{overhead}}}}{T_{2}^{\chi}}\) \cite{Barry2020} where $T_{2}^{\chi}$ is $T_2^\star$ for DC sensing, and $T_2$ for AC sensing, there are immediate opportunities to optimise the sensitivity. For example, the room-temperature optical contrast can be significantly improved by modest reductions in spin-lattice relaxation times; dephasing/coherence times can be improved through control of the nuclear spin bath\,\cite{Zadrozny2015}--e.g., through deuteration--or advanced control techniques\,\cite{Choi2020}; and overhead times can be significantly reduced through modified molecules with faster depopulation rates\,\cite{Ng2023}, which can be complemented with multi-level control. Likewise, the trade-off between spin concentration and electron-spin limited coherence time can be carefully optimised to maximise the sensitivity. Double-quantum protocols could further increase the sensitivity\,\cite{Barry2020}. Capitalising on the versatile fabrication techniques available to similar organic materials, the detection efficiency (hence \(n_\text{avg}\)) could also be significantly improved through photonic structures\,\cite{toninelli2021single}. Projecting that \(C=0.3\), \(c_S\simeq0.01\%\), \(T_2^{\star}=1\,\mu\text{s}\)\,\cite{Yang2000}, $t_{\text{overhead}}\simeq10\,\mu\text{s}$, $n_\text{avg}=0.01$ \cite{toninelli2021single} could be achieved with future optimisations, we find a volume-normalised sensitivity \(\eta_{\text{DC}}^V\simeq3\,\text{nT}\,\mu\text{m}^{3/2}\,\text{Hz}^{-1/2}\). Similarly, projecting \(T_2=4\,\mu\text{s}\)\,\cite{Sloop1981}, we find \(\eta_{\text{AC}}^V\simeq1\,\text{nT}\,\mathrm{\mu\text{m}^{3/2}}\,\text{Hz}^{-1/2}\). Interestingly, these sensitivities exceed the current state-of-the art values for solid-state defects \cite{Zhou2020}. Finally, and importantly, as highlighted above, sensing signal-to-noise ratios depend not only on the sensitivity, but also on minimising the proximity between sensor and target (which determines the sensing field) \cite{schaffry2011proposed}, something which molecular systems could be particularly well-suited for \cite{mullin2023quantum}. 

\section*{Outlook}\label{outlook}
Our demonstration of room-temperature optically detected coherent control of molecular spins opens up opportunities such as high density ensemble-based sensing with $>$100\,ppm active spin concentrations while maintaining coherence due to the well-defined chemistry of molecular materials. By harnessing chemical tunability, coupling to other physical quantities, such as electric fields \cite{Liu2021_elec}, could be further optimised. The bright optical transitions ($\sim$10\,ns fluorescence lifetime) of organic chromophores such as pentacene, combined with our demonstration of contrasts exceeding 10\%, indicates promise for room-temperature optically detected coherent control of individual chromophores. Additionally, organic triplets offer a complementary approach to interfacing nuclear spin memories under ambient conditions  \cite{Lee2013_ancilla}, combining optically detected coherent control of metastable states with deterministic placement of nuclear spins via synthetic chemistry \cite{Filidou2012,jackson2019nuclear}. Overall, our results indicate promise for chemically tunable room-temperature quantum technologies.

\section*{Methods}\label{methods}
\textbf{Crystal Growth.} Crystals of pentacene-doped para-terphenyl (0.01\%) were grown using the Bridgman method \cite{Cui2020}. Para-terphenyl (Sigma-Aldrich, \(\geq\)\,99.5\%) was extensively purified by zone-refining prior to use and ground into a fine powder with pentacene (99.9\%, Sigma-Aldrich) using a pestle and mortar. The powder was loaded into a borosilicate tube (6\,mm ID, Hilgenberg) with one end sealed into a fine point using a flame torch, before being sealed under argon inside a larger borosilicate tube (8\,mm ID, Simax). Bridgman growth was then performed at 3\,mm/hr over 3 days at 218 Celsius resulting in clear pink crystals. Finally, these crystals were attached to a glass slide using paraffin wax before being polished into plates (0.5\,mm thick) using silicon carbide paper loaded onto an EcoMet 30. Plate thickness was controlled using 0.5\,mm silicon plates as guides.
\newline\noindent \\
\textbf{Thin-film growth.} Pentacene-doped para-terphenyl thin films (100\,nm, 0.1\% mol/mol) were grown onto glass substrates using organic molecular beam deposition with para-terphenyl deposited at a rate of 5\,\AA/s, and the pentacene rate adjusted to achieve the desired concentration. The evaporation chamber was a Kurt J. Lesker Spectros system with a base pressure of $5\times10^{-7}\,\mathrm{mbar}$, and the rate was controlled automatically using quartz crystal microbalances positioned close to the sources and near the substrates. Glass plates were cleaned prior to use by 10 minutes of sonication in acetone and isopropanol, followed by drying under a stream of nitrogen and exposure to ozone for a further 10 minutes.
\newline\noindent \\
\textbf{cw-ODMR.} For cw-ODMR, the sample was excited using a \(520\)\,nm laser diode (ThorLabs LP520-SF15A), which was filtered using a \(550\)\,nm short-pass filter (ThorLabs FESH0550). A dichroic mirror (ThorLabs DMLP550R) was used to direct the excitation laser to a 1" achromatic doublet lens (ThorLabs AC254-030-AB) to focus the light on the sample. Photoluminescence was collimated by the same lens, separated from the excitation through the dichroic, and passed through a 600\,nm long-pass filter (ThorLabs FELH0600) and focussed using a second achromatic doublet (ThorLabs AC254-030-AB) onto a free-space detector (FEMTO, OE-200-SI) to detect the emission. The microwave fields were generated using a PCB co-planar waveguide (CPW) connected to a signal generator (Stanford Research Systems SG396) which produced square-wave modulated microwaves (117 Hz modulation rate). The PL signal from the free-space detector was fed into a lock-in amplifier (Stanford Research Systems, SR830) which was referenced to the microwave modulation frequency. The lock-in signal yielded the ODMR signal $\Delta$PL=PL(Microwaves on) - PL(Microwaves off). We normalise this by simultaneously measuring the DC PL on a data acquisition card (National Instruments PCIe-6363). To apply a variable magnetic field \(B_0\) we used a neodymium permanent magnet (Magnosphere, 1109) mounted on a motorised linear stage (Hephaist MX100ST100J, controlled by Thorlabs BSC203). The magnet was translated relative to the crystal with the field orientation perpendicular to the optical axis. The magnetic field strength was calibrated with a Hall sensor (Infineon TLE493D-P2B6MS2GO). See the Supplementary Figure S3 for a detailed diagram of the experiment. 
\newline\noindent \\
\textbf{Optically Detected Coherent Control}. Microwave pulses were gated using a high isolation microwave switch (Minicircuits ZASWA-2-50DRA+), with phase control from the in-built IQ modulation of the signal generator.  Microwave pulses were amplified (Minicircuits ZHL-25W-272+) before being sent to the PCB. To maximise microwave drive strength we used a short-terminated CPW. In the optical path, we replaced the 1" achromatic doublet lenses used for cw-ODMR with aspheric lenses: one to focus light onto the sample (ThorLabs C390TME-B), and one (Thorlabs A260-TM-B) to direct the emission into a single-mode fibre. The fibre was coupled to a single photon counting avalanche photodiode (APD, Excelitas SPCM-AQRH-14-FC) for PL detection. Voltage pulses from the APD were counted using a time-to-digital converter (Swabian Instruments TimeTagger 20). We drove the laser diode through a pulse generator (Agilent 8114A). Microwave and laser pulses, IQ modulation and photon counting were synchronised by an arbitrary waveform generator (Swabian Instruments Pulse Streamer 8/2). All pulsed measurements used a shot-to-shot  differential technique (see Supplementary Text S6 for full pulse sequences). For multi-level driving, we combined the output of a second signal generator (R\&S SMA100B) using a power splitter (Minicircuits ZFSC-2-372-S+), with both microwave tones gated by separate microwave switches, before amplification. See Supplementary Fig. S3 for a detailed diagram of the experiment.
\newline\noindent \\
\textbf{Sensitivity calculations}
The volume-normalised sensitivity was calculated using:
\(\eta^V \simeq \alpha\frac{\hbar e}{g_e \mu_B} \frac{1}{C\sqrt{\rho_S n_\text{avg}}} \frac{\sqrt{t_\text{overhead}}}{T_2^\chi}\)\,\cite{Barry2020}, where \(\alpha\) is 1 for the DC sensitivity and \(\pi/2\) for the AC sensitivity, \(g=2\) is the electronic \(g\) factor, \(\mu_\text{B}\) is the Bohr magneton, \(\hbar\) is the reduced Planck constant, \(\rho_S\) is the pentacene (i.e., spin) density, \(n_\text{avg}\) the average number of photons collected per spin per readout, and \(t_{\text{overhead}}\) the overhead time of the measurement (which accounts for initialisation, readout, and reset). We calculate the spin density from \(\rho_S = \frac{Z_\text{cell}}{V_\text{cell}} c_S\) where \(Z_\text{cell} = 2\) is the number of molecules in the p-terphenyl unit cell, $V_\text{cell} = 617\,\text{\AA}^3$\,\cite{Rice2013} is the p-terphenyl unit cell volume, and $c_S$ the pentacene doping concentration (mol/mol).
\\
\newline\noindent
\textbf{Supplementary information}
Supplementary information is available at [URL].
\newline\noindent \\
\textbf{Acknowledgments}
This work was supported by UK Research and Innovation [grant number MR/W006928/1] and the UK Engineering and Physical Sciences Research Council [grant numbers EP/W027542/1, EP/F039948/1, EP/F04139X/1, and EP/V048430/1]. A.M and D.R.M are supported by the Australian Research Council via the ARC Centre of Excellence in Exciton Science [grant number CE170100026]. A.M. was supported by the Sydney Quantum Academy, Sydney, NSW, Australia. E.B. was supported through the EPSRC and SFI Centre for Doctoral Training in the Advanced Characterisation of Materials (CDT-ACM) (grant number EP/S023259/1). We thank M. Oxborrow, A. Bowen, and W. Peveler for helpful discussions, and D. J. Paul, R. A. Hogg, and A. E. Kelly for experimental support.
\newline\noindent \\
\textbf{Author contributions}
A.M., S.K.M., and A.C. performed ODMR measurements. A.M., S.K.M., A.C., and S.L.B. analysed the ODMR data. M.A. prepared and characterised the crystal samples. M.A. and E.B. prepared and characterised the film samples. S.H., D.R.M., M.A., and S.L.B. provided oversight and supervision. A.M., S.K.M., A.C., and S.L.B. wrote the manuscript with input from all authors.
\newline\noindent \\
\textbf{Competing interests}
The authors declare no competing interests.
\newline\noindent \\
\textbf{Data availability}
The data underlying this work are available at [URL].

\end{document}